\documentclass[conference]{IEEEtran}
\IEEEoverridecommandlockouts
\usepackage{cite}
\usepackage{amsmath,amssymb,amsfonts}
\usepackage{algorithmic}
\usepackage{graphicx}
\usepackage{textcomp}
\usepackage{xcolor}
\usepackage{subfig}
\usepackage{booktabs}
\usepackage{import}
\usepackage{color}
\usepackage[citebordercolor=blue]{hyperref}
\usepackage[capitalise]{cleveref}
\usepackage{multirow}
\usepackage{array}
\usepackage[font=footnotesize]{caption}
\usepackage[inline]{enumitem}
\usepackage{makecell}
\usepackage{siunitx}
\usepackage{caption}
\usepackage{float}
\usepackage{soul}
\renewcommand\hl[1]{#1} 
\begin{document}
\title{Toward Deployable Satellite Anomaly Detection:\\A Benchmark Study on Large-Scale ESA-ADB Telemetry}
%
%
%

\author{\IEEEauthorblockN{
    Andrea~Nguyen\IEEEauthorrefmark{2}, Dafne~Rozenberg\IEEEauthorrefmark{2}, Yeying Zhu\IEEEauthorrefmark{3},~and~Peng~Hu\IEEEauthorrefmark{4}\textsuperscript{*}
\IEEEauthorblockA{
    \IEEEauthorrefmark{2}Faculty of Science, University of Manitoba, Winnipeg, Canada}
\IEEEauthorblockA{
    \IEEEauthorrefmark{3}Dept. of Statistics and Actuarial Science, University of Waterloo, Waterloo, Canada}
\IEEEauthorblockA{
    \IEEEauthorrefmark{4}Dept. of Electrical and Computer Engineering, University of Manitoba, Winnipeg, Canada}
    \{nguye75, rozenbed\}@myumanitoba.ca, yeying.zhu@uwaterloo.ca, *{peng.hu@umanitoba.ca}
}

\thanks{We acknowledge the support provided by the Natural Sciences and Engineering Research Council of Canada (NSERC), [funding reference number RGPIN-2022-03364].}
}

\maketitle

\begin{abstract}
Satellite anomaly detection is essential for maintaining mission reliability and spacecraft health, yet remains challenging due to the high-dimensional, irregular, and imbalanced nature of spacecraft telemetry data. This paper presents a systematic benchmark study evaluating supervised and unsupervised anomaly detection approaches on the large-scale ESA-ADB dataset across two mission settings of varying temporal scales. Supervised models, including Multiscale Convolutional Neural Networks (Multiscale CNN), Graph Convolutional Networks (GCN), and Graph Attention Networks (GAT), are compared against unsupervised methods, namely Elliptic Envelope (EE) and Empirical Cumulative Distribution Function-based Outlier Detection (ECOD). Beyond detection performance, we rigorously analyze computational runtime and scalability, which are critical for practical deployment in spacecraft operations. Results show that supervised models achieve stronger overall performance, while unsupervised methods offer competitive precision with significantly lower computational overhead. These findings underscore a fundamental trade-off between detection capacity and operational efficiency, offering practical guidance for mission engineers designing scalable satellite health monitoring systems.

\end{abstract}

\section{Introduction}

Spacecraft are complex and expensive systems consisting of thousands of telemetry channels that continuously monitor variables such as temperature, radiation levels, power consumption, instrumentation status, and computational activity \cite{Hundman_2018}. As space missions increase in complexity and scale, monitoring and maintaining spacecraft health has become increasingly challenging for spacecraft operations engineers. Detecting anomalies in satellite telemetry time-series data is therefore essential for ensuring the safe, reliable, and continuous operation of scientific, communication, observation, and navigation satellites. 

The primary goal of anomaly detection (AD) is to improve automated health monitoring systems by identifying unusual behavior, system faults, and rare events within large-scale multivariate telemetry streams. However, anomaly detection in spacecraft telemetry remains difficult because the data are high-dimensional, noisy, irregular, and highly imbalanced \cite{esaadb}. In practice, identifying anomalous events can be time-consuming and dependent on human expertise \cite{ruszczak2024opssatbenchmarkdetectinganomalies}. As a result, many spacecraft monitoring systems still rely on threshold-based alarm mechanisms that trigger when telemetry values exceed predefined limits. Failure to detect critical anomalies in time may result in partial or complete spacecraft loss \cite{Hundman_2018}.  

Machine learning (ML) approaches have shown strong potential for improving anomaly detection in satellite telemetry \cite{esaadb}. In \cite{esaadb}, the authors leverage the newly released ESA-ADB benchmark to evaluate supervised and unsupervised anomaly detection approaches on two mission datasets of varying temporal scales: ESA-Mission 1 (84 months) and ESA-Mission 2 (21 months). For supervised learning, three deep learning (DL) architectures are investigated: Multiscale Convolutional Neural Networks (Multiscale CNN), Graph Convolutional Networks (GCN), and Graph Attention Networks (GAT). For unsupervised learning, Empirical Cumulative Distribution-based Outlier Detection (ECOD) and Elliptic Envelope (EE) are evaluated as approaches capable of detecting anomalies without labeled data.

In this work, we present a systematic benchmarking study of anomaly detection for spacecraft telemetry, evaluating both supervised and unsupervised approaches on the recent comprehensive ESA-ADB dataset using state-of-the-art models from the literature. Our key contributions are threefold: (1) we provide the first comprehensive cross-paradigm comparison of Multiscale CNN, GCN, GAT, ECOD, and Elliptic Envelope under unified experimental conditions across multiple mission settings; (2) we go beyond detection accuracy by rigorously analyzing computational runtime and scalability—factors that are often overlooked yet critical for onboard or ground-segment deployment in real spacecraft operations; and (3) we derive practical guidelines for model selection that balance detection performance against operational constraints, offering actionable insights for mission engineers and the broader space anomaly detection community.

\section{Related Work}

Anomaly detection in satellite telemetry time-series data has become an important research area due to the increasing complexity of modern spacecraft systems. Recent advances in ML have significantly improved time-series anomaly detection performance. Hundman \textit{et al.} \cite{Hundman_2018} proposed one of the most widely used DL frameworks for spacecraft telemetry anomaly detection using Long Short-Term Memory (LSTM) networks combined with nonparametric dynamic thresholding. Their work demonstrated the effectiveness of deep sequential models for identifying anomalous spacecraft behavior in NASA telemetry datasets.

Beyond recurrent architectures, convolution-based approaches have shown strong performance for multivariate time-series analysis because they can efficiently capture temporal patterns across different resolutions and time scales. Multiscale CNN architectures are particularly effective for learning both short-term and long-term temporal dependencies in telemetry signals\hl{\mbox{\cite{elizar}}}. More recently, graph-based neural networks such as GCN and GAT have gained attention for anomaly detection tasks involving structured and interconnected data. These architectures are especially promising for spacecraft telemetry because satellite subsystems often exhibit strong inter-channel dependencies.

In addition to supervised DL approaches, unsupervised anomaly detection methods remain highly relevant because labeled anomaly data are limited in real-world spacecraft operations. Statistical and distribution-based methods are simple, computational efficient, and are able to operate without labeled data. Empirical Cumulative Distribution-based Outlier Detection (ECOD) and Elliptic Envelope are particularly suitable for highly imbalanced telemetry datasets where anomalous events are rare.

\section{Data Cleaning and Exploration}
\subsection{Data Preprocessing - Supervised models}
Raw satellite telemetry consists of highly asynchronous and non-uniformly sampled time series. Since the DL architectures evaluated in this work require uniformly sampled input matrices, a preprocessing pipeline was developed to transform the ESA-ADB telemetry into synchronized and memory-efficient datasets \cite{esaadb}.  

\textbf{Chronological Partitioning}: Following the ESA-ADB benchmark methodology \cite{esaadb}, the dataset was divided into predefined temporal windows. ESA-Mission 1 used training windows of 3, 10, 21, 42, and 84 months, while ESA-Mission 2 used 1, 5, 10, and 21 months.  

\textbf{Synchronization}: All timestamps were normalized to UTC time to ensure accurate alignment between telemetry signals and anomaly labels.  

\textbf{Global Grid Construction}: A uniform timeline was generated spanning the full temporal range of each mission split. A sampling interval of 30 seconds was selected to preserve short-duration anomalies while maintaining computational feasibility for the large-scale datasets.  

\textbf{Labeling}: Anomalies were mapped directly to telemetry streams using the ESA-ADB multi-class labeling scheme, where 0 represents nominal operation, 1 standard anomalies, 2 rare events, and 3 system-defined conditions \cite{esaadb}.  

\textbf{Temporal Resampling}: Zero-order hold (ZOH) interpolation was applied during resampling because it preserves causal relationships by preventing future information leakage into previous timestamps \cite{yang2018improvingclosedlooptrackingperformance}. This property is important for realistic real-time anomaly detection scenarios.  

\textbf{Memory Optimization}: Due to the scale of ESA-Mission 1 and ESA-Mission 2, feature vectors were downcast to 32-bit floating-point precision and labels to 8-bit integers. In addition, intermediate datasets were incrementally merged and removed from active memory to reduce memory consumption during preprocessing.  

\subsection{Data Preprocessing - Unsupervised models}
A similar preprocessing strategy was applied for the unsupervised anomaly detection models to ensure consistency with the ESA-ADB benchmark \cite{esaadb}. Telemetry channels stored in compressed \texttt{.zip} files were extracted into pandas DataFrames, converted to datetime format, and sorted chronologically. Anomaly labels from the provided \texttt{labels.csv} file were aligned with telemetry timestamps after removing inconsistent time zone information.  

Because the raw telemetry was irregularly sampled, the data were resampled at fixed 10-second intervals using zero-order hold interpolation. The dataset was then divided chronologically into training, validation, and test sets to prevent future information leakage. The first half of the time series was used for training and validation, while the second half was reserved for testing. Within the training portion, the final three months were allocated for validation.  

Finally, Min-Max normalization was applied using statistics computed exclusively from the training data before being applied to the validation and test sets. Ground-truth anomaly labels were converted into aligned binary time-series labels, where 1 represents anomalous behavior and 0 represents nominal operation.

\section{Proposed Methodology}
\subsection{Supervised Models}
\textbf{Experimental Setup and Hyperparameters}

The supervised models were evaluated under a standardized training configuration to ensure consistent comparison across the ESA-ADB Mission 1 and Mission 2 datasets. All models were optimized using the Adam optimizer with a learning rate of 0.001 and a batch size of 1024. Since anomaly detection is formulated as a binary classification task, the models were trained using Binary Cross-Entropy (BCE) loss for 10 epochs.  

To address the severe class imbalance in satellite telemetry, where nominal events greatly outnumber anomalies, weight decay ($1\times10^{-5}$) was incorporated into the GAT optimizer to reduce overfitting to the dominant class and preserve sensitivity to rare anomalies. The primary configuration difference across models was the input sequence length ($L$), selected to balance temporal feature extraction and computational cost. The hyperparameter configurations are summarized in Table~\ref{tab:hyperparameter}.  

\begin{table}[H]
\centering
\caption{Hyperparameter Configurations by Model and Mission Dataset}
\resizebox{\columnwidth}{!}{%
\begin{tabular}{llccccc}
\toprule
\textbf{Model} & \textbf{Dataset} & \textbf{Data Window} & \textbf{Batch Size} & \textbf{Seq. Length} & \textbf{Learning Rate} & \textbf{Epochs} \\ 
\midrule
\multirow{2}{*}{Multiscale CNN} 
& Mission 1 & 84 months & 1024 & 100  & 0.001 & 10 \\ 
& Mission 2 & 21 months & 1024 & 100 & 0.001 & 10 \\ 
\midrule
\multirow{2}{*}{GCN} 
& Mission 1 & 84 months & 1024  & 15  & 0.001 & 10 \\ 
& Mission 2 & 21 months & 1024 & 15  & 0.001 & 10 \\ 
\midrule
\multirow{2}{*}{GAT} 
& Mission 1 & 84 months & 1024 & 50 & 0.001 & 10 \\ 
& Mission 2 & 21 months & 1024 & 50  & 0.001 & 10 \\ 
\bottomrule
\end{tabular}%
} 
\label{tab:hyperparameter}
\end{table}

The sequence lengths reflect the architectural constraints of each model. The Multiscale CNN used the largest sequence length ($L=100$) to capture long-range temporal patterns across multiple scales. The GCN used a smaller window ($L=15$) because graph convolutions with static adjacency matrices become computationally expensive at larger temporal depths. The GAT used an intermediate sequence length ($L=50$), providing sufficient temporal context for the attention mechanism while maintaining computational feasibility. Mission duration also influenced runtime, with ESA-Mission 1 containing 84 months of telemetry and ESA-Mission 2 containing 21 months.  

\subsubsection{Multiscale Convolutional Neural Networks (Multiscale CNN)}
Conventional CNN-based architectures often suffer from information loss and limited feature representation when processing multivariate time-series data \cite{elizar}. To better capture telemetry patterns across multiple temporal resolutions, a multiscale CNN architecture with three parallel convolutional branches was implemented.  

The input tensor $X$ with sequence length $L=100$ is processed through parallel 1D convolutional layers with kernel sizes $k=3,5,7$, enabling extraction of short-, medium-, and long-term temporal patterns. Each branch applies 64 filters followed by a Rectified Linear Unit (ReLU) activation and max-pooling:
\[
H_k = \text{MaxPool}(\text{ReLU}(\text{Conv1D}_k(X)))
\]
The resulting feature maps are concatenated, flattened, and passed through a multi-layer perceptron (MLP). A dropout layer ($p=0.3$) is applied before the final Sigmoid classification layer to reduce overfitting and generate anomaly probabilities.  

\subsubsection{Graph Convolutional Network (GCN)} 
The GCN extends convolutional learning to graph-structured data, enabling relationships between telemetry channels to be modeled explicitly \cite{jiang}. A static undirected graph $G=(V,E)$ was constructed where each node represents a telemetry channel. Edges were defined using the absolute Pearson correlation matrix of the training data, with an edge added when the correlation exceeded a threshold of $\tau=0.8$. Node features consisted of raw temporal histories with sequence length $L=15$.  

The model used a two-layer graph convolution mechanism. Let $\tilde{A}=A+I_N$ denote the adjacency matrix with self-loops and $\tilde{D}$ the associated degree matrix. The propagation rule is:

\[
H^{(l+1)} = \text{ReLU}\left(\tilde{D}^{-\frac{1}{2}} \tilde{A} \tilde{D}^{-\frac{1}{2}} H^{(l)} W^{(l)}\right)
\]

The network projected temporal features into a 32-dimensional hidden representation before aggregation into a 16-dimensional embedding space. The embeddings were flattened and passed through a dense layer for binary anomaly classification.  

\subsubsection{Graph Attention Network (GAT)} 

Unlike GCNs, which assume fixed relationships between telemetry channels, GATs dynamically learn the importance of neighboring nodes through attention mechanisms \cite{velickovic}. This is particularly useful for spacecraft telemetry because subsystem interactions may vary over time.  

The graph topology was initialized using Pearson correlation with a relaxed threshold ($\tau=0.5$) to allow broader connectivity. Temporal sequences of length $L=50$ were processed using a custom multi-head attention layer. For each node pair $(i,j)$, the attention coefficient is computed as:

\[
e_{ij} = \text{LeakyReLU}\left(\mathbf{a}^T [W\mathbf{h}_i \, \Vert \, W\mathbf{h}_j]\right)
\]
where $W$ is a learnable linear transformation and $\mathbf{a}$ contains the attention parameters. Attention values were masked using the graph topology so that only connected nodes contributed to the aggregation process.  

The first GAT layer used four attention heads with hidden dimension 64, while the second layer used a single attention head to project features into a 32-dimensional representation. Exponential Linear Unit (ELU) activations and dropout regularization were applied before the final MLP anomaly classifier.  

\subsection{Unsupervised Models}
\subsubsection{Elliptic Envelope}
Elliptic Envelope is a statistical anomaly detection method that assumes nominal data follow a multivariate Gaussian distribution \cite{ee}. The method estimates the central data distribution using the Minimum Covariance Determinant estimator and identifies anomalies based on Mahalanobis distance:

\[
d(\mathbf{x}) = \sqrt{(\mathbf{x} - \boldsymbol{\mu})^T \boldsymbol{\Sigma}^{-1} (\mathbf{x} - \boldsymbol{\mu})}
\]

Samples exceeding a threshold $\tau$ are classified as anomalies:

\[
f(\mathbf{x}) =
\begin{cases}
1, & \text{if } d(\mathbf{x}) > \tau \\
0, & \text{otherwise}
\end{cases}
\]

The contamination parameter was set to 0.001 to reflect the rarity of anomalies in spacecraft telemetry, while the support fraction was fixed at 0.9 to improve robustness against outliers. Since the model outputs continuous anomaly scores, the decision threshold was selected from the validation set using the $99.9^{th}$ percentile of anomaly scores.  

\subsubsection{ECOD (Empirical Cumulative Distribution Function-based Outlier Detection)}
ECOD is a non-parametric anomaly detection method that identifies outliers based on empirical cumulative distribution functions (ECDFs) without assuming any underlying data distribution \cite{ecod}. For each feature $j$, the ECDF is defined as:

\[
F_j(x) = \frac{1}{n} \sum_{i=1}^{n} \mathbf{1}(x_i^{(j)} \leq x)
\]

Lower and upper tail probabilities are computed as:

\[
p_j^{\text{lower}}(x) = F_j(x), \quad p_j^{\text{upper}}(x) = 1 - F_j(x)
\]

The feature-wise tail probability is:

\[
p_j(x) = \min \left( p_j^{\text{lower}}(x), \, p_j^{\text{upper}}(x) \right)
\]

The final anomaly score is calculated as:
\[
\text{Score}(\mathbf{x}) = - \sum_{j=1}^{d} \log \left( p_j(x) \right)
\]

Points located in extreme regions of the feature distributions receive higher anomaly scores. ECOD is computationally efficient and well suited for high-dimensional telemetry data because it avoids distributional assumptions and extensive parameter tuning. Similar to Elliptic Envelope, binary predictions were generated using a threshold selected from the validation set at the $99.9^{th}$ percentile of anomaly scores.

\section{Experimental Results}
\subsection{Evaluation Metrics}
Model performance was evaluated using event-based metrics rather than point-wise metrics, since anomalies in telemetry data occur over time intervals instead of isolated timestamps. A predicted anomaly event is considered correct if it overlaps with a ground-truth anomaly interval. Let $TP_e$, $FP_e$, and $FN_e$ denote the number of true positive, false positive, and false negative events, respectively.  

Precision measures the proportion of predicted anomaly events that are correct:
$Pr_e = \frac{TP_e}{TP_e + FP_e}$.

Recall measures the proportion of true anomaly events successfully detected: $Rec_e = \frac{TP_e}{TP_e+FN_e}$.

To balance precision and recall, the $F_{0.5}$-score was used:
\[F_{0.5} = (1+0.5^2)*\frac{Pr_e*Rec_e}{(0.5^2*Pr_e)+Rec_e}\]

A weighting factor of 0.5 emphasizes precision over recall, reflecting the operational importance of reducing false alarms in spacecraft monitoring systems, where unnecessary investigations can consume substantial engineering resources.  

Event-based accuracy was also computed:

\[AC_e= \frac{TP_e}{TP_e+FP_e+FN_e }\]

Finally, the Precision-Recall Area Under the Curve (PR-AUC) was used to evaluate model performance across varying decision thresholds:

\[PR-AUC = \int_{0}^{1}{Pr_e(r)dr}\]

where $r$ denotes recall. PR-AUC is particularly suitable for highly imbalanced anomaly detection problems because it captures the trade-off between precision and recall over the full range of anomaly scores.  

Together, these metrics provide a comprehensive evaluation of anomaly detection performance while emphasizing robustness in highly imbalanced spacecraft telemetry datasets.

\subsection{Model Performance}

\begin{table}[h]
\centering
\caption{Mission 1 – trained and tested on the lightweight subset of channels 41-46}
\label{table:mission1_lightweight}
\resizebox{\linewidth}{!}{
\begin{tabular}{lccccc}
\toprule
\textbf{Model} & \textbf{Precision} & \textbf{Recall} & \textbf{$F_{0.5}$} & \textbf{Accuracy} & \textbf{PR-AUC} \\ 
\midrule
Multiscale CNN      & 0.5036 & 0.2321 & 0.4081 & 0.8874 & 0.3313 \\ 
GCN                 & 0.1709 & 0.1472 & 0.1656 & 0.7659 & 0.1777 \\ 
GAT                 & 0.8435 & 0.0115 & 0.0545 & 0.8981 & 0.1121 \\ 
ECOD                & 0.6017 & 0.2496 & 0.4484 & 0.2056 & 0.5264\\
Elliptic Envelope   & 0.002 & 0.4634 & 0.0026 & 0.002 & 0.4636\\

\bottomrule
\end{tabular}
}
\label{tab:small_m1}
\end{table}

The models were first evaluated on a lightweight subset of ESA-Mission 1 containing Channels 41-46. The evaluation metrics are summarized in Table~\ref{tab:small_m1}. As shown in Table~\ref{table:mission1_lightweight}, both supervised and unsupervised approaches demonstrated distinct strengths on the lightweight Mission 1 subset. Among the supervised models, the Multiscale CNN achieved the strongest balance between precision and anomaly sensitivity, obtaining the highest PR-AUC (0.3313) and an $F_{0.5}$ score of 0.4081. In contrast, the GAT achieved the highest precision (0.8435) and accuracy (0.8981), but its extremely low recall (0.0115) indicates that many anomaly events were missed.  

Among the unsupervised methods, ECOD achieved competitive performance, obtaining a precision of 0.6017 and the highest overall $F_{0.5}$ score (0.4484). Compared to ESA-ADB baseline results, where several methods achieved near-zero precision, ECOD maintained a stronger precision-recall balance while still detecting a reasonable proportion of anomalies. Although Elliptic Envelope achieved higher recall (0.4634), its extremely low precision (0.002) indicates a large number of false positive alarms.  

The transition from the lightweight subset to the full Mission 1 channel set resulted in notable performance changes, as shown in Table~\ref{tab:full_m1}.  

\begin{table}[h]
\centering
\caption{Mission 1 – trained and tested on the full set of channels}
\resizebox{\linewidth}{!}{
\label{table:mission1_full}
\begin{tabular}{lccccc}
\toprule
\textbf{Model} & \textbf{Precision} & \textbf{Recall} & \textbf{$F_{0.5}$} & \textbf{Accuracy} & \textbf{PR-AUC} \\ 
\midrule
Multiscale CNN      & 0.6166 & 0.5306 & 0.5972 & 0.9626 & 0.4674 \\ 
GCN                 & 0.5709 & 0.6128 & 0.5788 & 0.9603 & 0.4414 \\ 
GAT                 & 0.8318 & 0.6272 & 0.7809 & 0.9766 & 0.7374 \\ 
ECOD                & 0.0765 & 0.4325 & 0.065 & 0.043 & 0.2167\\
Elliptic Envelope   & 0.0069 & 0.4612 & 0.0078 & 0.006 & 0.2451\\
\bottomrule
\end{tabular}}
\label{tab:full_m1}
\end{table}

As shown in Table~\ref{table:mission1_full}, the supervised models improved substantially when trained on the full Mission 1 dataset. GAT achieved the strongest overall performance across all primary metrics, suggesting that its attention mechanism effectively captures complex inter-channel relationships when broader telemetry context is available.  

In contrast, the unsupervised methods experienced a noticeable decrease in performance on the full channel set. Nevertheless, ECOD maintained relatively stronger precision compared to many baseline methods reported in the ESA-ADB literature. Although its precision decreased to 0.0765, this remained higher than the near-zero precision reported for several baseline approaches. Elliptic Envelope achieved slightly higher recall but produced a large number of false positive predictions.  

The same models were evaluated on ESA-Mission 2 using a lightweight subset consisting of Channels 18--28. The corresponding results are summarized in Table~\ref{tab:small_m2}.

\begin{table}[h]
\centering
\caption{Mission 2 trained and tested on the lightweight subset of channels 18-28}
\label{table:mission2_lightweight}
\resizebox{\linewidth}{!}{
\begin{tabular}{lccccc}
\toprule
\textbf{Model} & \textbf{Precision} & \textbf{Recall} & \textbf{$F_{0.5}$} & \textbf{Accuracy} & \textbf{PR-AUC} \\ 
\midrule
Multiscale CNN      & 0.7794 & 0.5299 & 0.7123 & 0.9710 & 0.6235 \\ 
GCN                 & 0.5709 & 0.6128 & 0.5788 & 0.9603 & 0.4414 \\ 
GAT                 & 0.7726 & 0.2518 & 0.5465 & 0.9605 & 0.2311 \\ 
ECOD                & 0.7554 & 0.1982 & 0.2187 & 0.1014 & 0.4721\\
Elliptic Envelope   & 1 & 0.0665 & 0.1679 & 0.0665 & 0.5302\\
\bottomrule
\end{tabular}}
\label{tab:small_m2}
\end{table}

Table~\ref{table:mission2_lightweight} shows that the Multiscale CNN achieved the strongest supervised performance on the lightweight Mission 2 subset, obtaining the highest $F_{0.5}$ score (0.7123) and PR-AUC (0.6235). The GCN achieved the highest recall (0.6128), while the GAT maintained strong precision (0.7726) but lower PR-AUC performance.  

Among the unsupervised methods, ECOD again demonstrated competitive performance, achieving a precision of 0.7554 comparable to the best supervised models. Although its recall remained relatively low (0.1982), the high precision suggests that detected anomalies are highly reliable. Elliptic Envelope achieved perfect precision (1.0), but its extremely low recall (0.0665) indicates that only a very small subset of anomalies was detected.  
The transition from the lightweight subset to the full Mission 2 dataset produced significant changes in model performance, as summarized in Table~\ref{tab:full_m2}.

\begin{table}[h]
\centering
\caption{Mission 2 trained and tested on the full set of channels}
\label{table:mission2_full}
\resizebox{\linewidth}{!}{
\begin{tabular}{lccccc}
\toprule
\textbf{Model} & \textbf{Precision} & \textbf{Recall} & \textbf{$F_{0.5}$} & \textbf{Accuracy} & \textbf{PR-AUC} \\ 
\midrule
Multiscale CNN      & 0.2431 & 0.6871 & 0.2792 & 0.7278 & 0.3219 \\ 
GCN                 & 0.1676 & 0.7868 & 0.1989 & 0.5426 & 0.1890 \\ 
GAT                 & 0.1460 & 0.6213 & 0.1724 & 0.5548 & 0.1842 \\ 
ECOD                & 0.1882 & 0.2066 & 0.0619 & 0.0328 & 0.1208\\
Elliptic Envelope   & 0.2012 & 0.2877 & 0.0725 & 0.036 & 0.2174\\
\bottomrule
\end{tabular}}
\label{tab:full_m2}
\end{table}

For Mission 2 using the full telemetry dataset, the Multiscale CNN achieved the strongest overall performance, obtaining the highest PR-AUC (0.3219) and $F_{0.5}$ score (0.2792). Although its precision remained relatively low (0.2431), the model maintained a stronger balance between anomaly sensitivity and false alarm reduction compared to the other approaches. The GCN achieved the highest recall (0.7868), indicating strong anomaly detection capability.  

Both ECOD and Elliptic Envelope experienced performance degradation on the complete channel set, consistent with the increased difficulty and dimensionality of the dataset. Nevertheless, both methods maintained precision values between 0.18 and 0.20. Overall, the supervised DL models, particularly the Multiscale CNN, achieved stronger overall anomaly detection performance, while the unsupervised approaches provided simpler and computationally efficient alternatives.

\subsection{Computation Complexity and Runtime Analysis}
To evaluate computational efficiency, we measured runtime per epoch and total training time for each model across Mission 1 and Mission 2 datasets, as summarized in Tables~\ref{tab:cost_m1} and \ref{tab:cost_m2}.

\begin{table}[h]
\centering
\caption{Computational Runtime Evaluation for Mission 1}
\label{table:runtime_mission1}
\resizebox{0.8\linewidth}{!}{
\begin{tabular}{lcc}
\toprule
\textbf{Model} & \textbf{Runtime per Epoch} & \textbf{Total Runtime} \\ 
\midrule
Multiscale CNN & $\approx$0.5 hours & $\approx$7 hours \\ 
GCN & $\approx$9.0 hours & $\approx$91 hours \\ 
GAT & $\approx$5.0 hours & $\approx$51 hours \\ 
\bottomrule
\end{tabular}}
\label{tab:cost_m1}
\end{table}

\begin{table}[h]
\centering
\caption{Computational Runtime Evaluation for Mission 2}
\label{table:runtime_mission2}
\resizebox{0.8\linewidth}{!}{
\begin{tabular}{lcc}
\toprule
\textbf{Model} & \textbf{Runtime per Epoch} & \textbf{Total Runtime} \\ 
\midrule
Multiscale CNN & $\approx$0.5 hours & $\approx$5 hours \\ 
GCN & $\approx$0.25 hours & $\approx$3.5 hours \\ 
GAT & $\approx$1.0 hours & $\approx$11 hours \\ 
\bottomrule
\end{tabular}}
\label{tab:cost_m2}
\end{table}

For Mission 1, the Multiscale CNN required approximately 9 hours per epoch (91 hours total), while the GAT required 5 hours per epoch (51 hours total). The GCN was the most computationally efficient, requiring approximately 0.5 hours per epoch (7 hours total). This indicates that graph-based models exhibit substantially lower training cost, particularly when using simpler convolutional aggregation schemes.

For Mission 2, overall runtime decreased due to the reduced temporal history (21 vs. 84 months). The GCN remained the fastest model at approximately 0.25 hours per epoch (3.5 hours total), followed by the Multiscale CNN (0.5 hours per epoch, 5 hours total). The GAT also became more efficient (1 hour per epoch, 11 hours total), but remained the most expensive graph-based method.

Overall, the GCN consistently provides the lowest computational cost across both missions, making it suitable for rapid experimentation and large-scale telemetry processing. In contrast, the GAT is more sensitive to dataset size and complexity, as reflected by its significantly higher runtime in Mission 1.

For unsupervised methods, runtime is substantially lower since ECOD and Elliptic Envelope do not require iterative training. As shown in Table~\ref{tab:time}, both methods complete training and inference within approximately 0.5 hours on lightweight subsets, increasing to 1.5–2 hours on full datasets. ECOD shows slightly higher runtime on full data, reflecting its sensitivity to dimensionality, although the difference remains modest.

\begin{table}[H]
\centering
\caption{Computational Runtime Evaluation for each model}
\label{table:runtime_mission1}
\begin{tabular}{lcc}
\toprule
\textbf{Model} & \textbf{Lightweight channels} & \textbf{Full dataset} \\ 
\midrule
ECOD & $\approx$0.5 hours & $\approx$2 hours \\ 
Elliptic Envelope & $\approx$0.5 hours & $\approx$1.5 hours \\ 
\bottomrule
\end{tabular}
\label{tab:time}
\end{table}

Overall, these results highlight a clear trade-off between computational efficiency and model complexity: simpler statistical methods are computationally efficient but less expressive, while DL approaches provide stronger modeling capacity at significantly higher computational cost.

\section{Conclusive Remarks}
In this work, both supervised and unsupervised learning techniques for AD in satellite telemetry data were explored. 

For supervised models, results show that no single architecture is universally optimal; performance depends on dataset characteristics such as dimensionality and temporal complexity. On Mission 1, the Graph Attention Network (GAT) achieved the best overall performance, followed by the GCN and Multiscale CNN. This can be attributed to the GAT’s ability to model dynamic inter-channel dependencies through its attention mechanism, which is particularly effective in lower-dimensional settings.

However, this ranking changes in Mission 2. The Multiscale CNN becomes the most robust model, followed by the GCN and then the GAT. In higher-dimensional telemetry, the multiscale convolutional structure better captures patterns across different temporal resolutions, while the GAT appears more sensitive to increased noise and feature sparsity, leading to reduced performance.

Beyond accuracy, computational efficiency is a key consideration for deployment. The Multiscale CNN is the most efficient in Mission 1, while the GCN is the fastest in Mission 2. The GAT consistently incurs the highest computational cost. Overall, the Multiscale CNN provides the most consistent trade-off between performance and efficiency across both missions, despite the GAT achieving the highest $F_{0.5}$ and PR-AUC in Mission 1.

Unsupervised methods, particularly ECOD, also demonstrate competitive performance in terms of precision when compared to more complex models reported in the ESA-ADB literature. This is largely due to ECOD’s distribution-free formulation, which identifies anomalies based on extreme deviations rather than learned representations. This makes it particularly suitable for highly imbalanced settings where anomalous events are rare.

From a computational perspective, unsupervised methods are significantly more efficient, requiring only a few hours for full dataset processing. This makes them attractive for real-time or resource-constrained satellite monitoring scenarios. However, this efficiency comes with trade-offs: ECOD generally exhibits lower recall, and Elliptic Envelope is sensitive to distributional assumptions, leading to instability in high-dimensional settings.

Overall, the results highlight a clear trade-off between detection performance and computational efficiency. Supervised models achieve higher PR-AUC and $F_{0.5}$ scores but require substantially greater computational resources, while unsupervised methods offer lightweight and scalable alternatives with reduced detection sensitivity.

In practical satellite operations, the choice of method depends on operational constraints: supervised models are preferable when labeled data and computational resources are available, whereas unsupervised methods are better suited for rapid screening or scenarios with limited annotations.


%

\ifCLASSOPTIONcaptionsoff
  \newpage
\fi




\bibliographystyle{IEEEtran}
\bibliography{references}

\begin{thebibliography}{1}
\providecommand{\url}[1]{#1}
\csname url@samestyle\endcsname
\providecommand{\newblock}{\relax}
\providecommand{\bibinfo}[2]{#2}
\providecommand{\BIBentrySTDinterwordspacing}{\spaceskip=0pt\relax}
\providecommand{\BIBentryALTinterwordstretchfactor}{4}
\providecommand{\BIBentryALTinterwordspacing}{\spaceskip=\fontdimen2\font plus
\BIBentryALTinterwordstretchfactor\fontdimen3\font minus \fontdimen4\font\relax}
\providecommand{\BIBforeignlanguage}[2]{{%
\expandafter\ifx\csname l@#1\endcsname\relax
\typeout{** WARNING: IEEEtran.bst: No hyphenation pattern has been}%
\typeout{** loaded for the language `#1'. Using the pattern for}%
\typeout{** the default language instead.}%
\else
\language=\csname l@#1\endcsname
\fi
#2}}
\providecommand{\BIBdecl}{\relax}
\BIBdecl

\bibitem{Hundman_2018}
\BIBentryALTinterwordspacing
K.~Hundman, V.~Constantinou, C.~Laporte, I.~Colwell, and T.~Soderstrom, ``Detecting spacecraft anomalies using lstms and nonparametric dynamic thresholding,'' in \emph{Proceedings of the 24th ACM SIGKDD International Conference on Knowledge Discovery \&amp; Data Mining}, ser. KDD ’18.\hskip 1em plus 0.5em minus 0.4em\relax ACM, Jul. 2018, p. 387–395. [Online]. Available: \url{http://dx.doi.org/10.1145/3219819.3219845}
\BIBentrySTDinterwordspacing

\bibitem{esaadb}
\BIBentryALTinterwordspacing
K.~Kotowski, C.~Haskamp, J.~Andrzejewski, B.~Ruszczak, J.~Nalepa, D.~Lakey, P.~Collins, A.~Kolmas, M.~Bartesaghi, J.~Martinez-Heras, and G.~De~Canio, ``European space agency benchmark for anomaly detection in satellite telemetry,'' \emph{arXiv preprint arXiv:2406.17826}, 2024. [Online]. Available: \url{https://arxiv.org/abs/2406.17826}
\BIBentrySTDinterwordspacing

\bibitem{ruszczak2024opssatbenchmarkdetectinganomalies}
\BIBentryALTinterwordspacing
B.~Ruszczak, K.~Kotowski, D.~Evans, and J.~Nalepa, ``The ops-sat benchmark for detecting anomalies in satellite telemetry,'' 2024. [Online]. Available: \url{https://arxiv.org/abs/2407.04730}
\BIBentrySTDinterwordspacing

\bibitem{elizar}
E.~Elizar, M.~A. Zulkifley, R.~Muharar, M.~H. Mohd~Zaman, and S.~Mustaza, ``A review on multiscale-deep-learning applications,'' \emph{Sensors}, vol.~22, p. 7384, 09 2022.

\bibitem{yang2018improvingclosedlooptrackingperformance}
\BIBentryALTinterwordspacing
C.~Yang, S.~Gao, and Z.~Xue, ``Improving the closed-loop tracking performance using the first-order hold sensing technique with experiments,'' 2018. [Online]. Available: \url{https://arxiv.org/abs/1801.01263}
\BIBentrySTDinterwordspacing

\bibitem{jiang}
J.~Jiang, J.~Chen, T.~Gu, K.-K.~R. Choo, C.~Liu, M.~Yu, W.~Huang, and P.~Mohapatra, ``Anomaly detection with graph convolutional networks for insider threat and fraud detection,'' in \emph{MILCOM 2019 - 2019 IEEE Military Communications Conference (MILCOM)}, 2019, pp. 109--114.

\bibitem{velickovic}
\BIBentryALTinterwordspacing
P.~Veličković, G.~Cucurull, A.~Casanova, A.~Romero, P.~Liò, and Y.~Bengio, ``Graph attention networks,'' 2018. [Online]. Available: \url{https://arxiv.org/abs/1710.10903}
\BIBentrySTDinterwordspacing

\bibitem{ee}
\BIBentryALTinterwordspacing
E.~Chandralekha, S.~Vinodhini, V.~Kandasamy, and P.~Rama, ``Heart rate anomaly detection in healthcare using elliptic envelope and local forest,'' \emph{Procedia Computer Science}, vol. 258, pp. 1677--1687, 2025. [Online]. Available: \url{https://www.sciencedirect.com/science/article/pii/S1877050925015054}
\BIBentrySTDinterwordspacing

\bibitem{ecod}
\BIBentryALTinterwordspacing
Z.~Li, Y.~Zhao, N.~Botta, C.~Ionescu, and X.~Hu, ``Ecod: Unsupervised outlier detection using empirical cumulative distribution functions,'' \emph{IEEE Transactions on Knowledge and Data Engineering}, 2022. [Online]. Available: \url{https://arxiv.org/abs/2201.00382}
\BIBentrySTDinterwordspacing

\end{thebibliography}





%




\end{document}